\documentclass[twocolumn,trackchanges]{63}
\usepackage{silence} 
\usepackage{natbib}
\usepackage{amsmath}
\usepackage{float}
\usepackage{times}
\usepackage{enumitem}
\usepackage{etoolbox}
\usepackage{graphicx}
\usepackage{amsmath}
\usepackage{bm}
\usepackage{graphics}
\usepackage{url}
\usepackage{epsfig}
\usepackage{wrapfig}
\usepackage{ulem}
\usepackage{verbatim}
\usepackage{float}
\usepackage{lastpage}
\usepackage{CJK}
\usepackage[whole]{bxcjkjatype} 
\usepackage{CJKutf8}
\usepackage[caption=false]{subfig}

\def\lsim{\hbox{\rlap{\raise 0.425ex\hbox{$<$}}\lower 0.65ex\hbox{$\sim$}}}
\def\gsim{\hbox{\rlap{\raise 0.425ex\hbox{$>$}}\lower 0.65ex\hbox{$\sim$}}}

\def\arcdeg{\mbox{$^\circ$}}

\newcommand\N[1]{{\color{blue} \bf #1}} 

\shorttitle{Ultraviolet Spectroscopy of SN\,2022wsp}
\shortauthors{Vasylyev et al.}

\begin{document}

\title{Early-Time Ultraviolet and Optical {\it Hubble Space Telescope} Spectroscopy of the Type II Supernova 2022wsp}
\correspondingauthor{Sergiy~S.Vasylyev}
\author[0000-0002-4951-8762]{Sergiy~S.Vasylyev}
\email{sergiy\_vasylyev@berkeley.edu}
\affiliation{Department of Astronomy, University of California, Berkeley, CA 94720-3411, USA}
\affiliation{Steven Nelson Graduate Fellow}

\author[0000-0002-7941-5692]{Christian Vogl}
\affiliation{Max-Planck-Institut f\"ur Astrophysik, Karl-Schwarzschild-Str. 1, 85748 Garching, Germany}

\correspondingauthor{Yi Yang}
\author[0000-0002-6535-8500]{Yi Yang
\begin{CJK}{UTF8}{gbsn}
(杨轶)
\end{CJK}}\email{yiyangtamu@gmail.com}
\affiliation{Department of Astronomy, University of California, Berkeley, CA 94720-3411, USA}
\affiliation{Bengier-Winslow-Robertson Postdoctoral Fellow}

\author[0000-0003-3460-0103]{Alexei~V.~Filippenko}
\affiliation{Department of Astronomy, University of California, Berkeley, CA 94720-3411, USA}

\author{Thomas~G.~Brink}
\affiliation{Department of Astronomy, University of California, Berkeley, CA 94720-3411, USA}
\affiliation{Wood Specialist in Astronomy}
\author[0000-0001-6272-5507]{Peter J. Brown}
\affiliation{George P. and Cynthia Woods Mitchell Institute for Fundamental Physics and Astronomy, and \\ 
Department of Physics and Astronomy, Texas A\&M University, 4242 TAMU, College Station, TX 78712, USA}

\author[0000-0001-6685-0479]{Thomas Matheson}
\affiliation{NSF's National Optical-Infrared Astronomy Research Laboratory\\
950 North Cherry Avenue\\
Tucson, AZ 85719, USA}

\author[0000-0002-3653-5598]{Avishay Gal-Yam}
\affiliation{Department of Particle Physics and Astrophysics, Weizmann Institute of Science, 234 Herzl St., 76100 Rehovot, Israel}
\author{Paolo A. Mazzali}
\affiliation{Max-Planck-Institut f\"{u}r Astrophysik, Karl-Schwarzschild-Str. 1, 85748 Garching, Germany}

\author[0000-0001-6069-1139]{Thomas de Jaeger}
\affiliation{LPNHE, CNRS/IN2P3 \& Sorbonne Universit\'e, 4 place Jussieu, 75005 Paris, France}

\author[0000-0002-1092-6806]{Kishore C. Patra}
\affiliation{Department of Astronomy, University of California, Berkeley, CA 94720-3411, USA}
\affiliation{Nagaraj-Noll-Otellini Graduate Fellow}

\author[0009-0000-2503-140X]{Gabrielle E. Stewart}
\affiliation{Department of Astronomy, University of California, Berkeley, CA 94720-3411, USA}

\begin{abstract}
We report early-time ultraviolet (UV) and optical spectroscopy of the young, nearby Type II supernova (SN) 2022wsp obtained by the {\it Hubble Space Telescope (HST)}/STIS at about 10 and 20 days after the explosion.
The SN\,2022wsp UV spectra are compared to those of other well-observed Type II/IIP SNe, including the recently studied Type IIP SN\,2021yja. Both SNe exhibit rapid cooling and similar evolution during early phases, indicating a common behavior among SNe~II. 
Radiative-transfer modeling of the spectra of SN\,2022wsp with the \texttt{TARDIS} code indicates a steep radial density profile in the outer layer of the ejecta, a supersolar metallicity, and a relatively high total extinction of $E(B-V) = 0.35$\,mag. The early-time evolution of the photospheric velocity and temperature derived from the modeling agree with the behavior observed from other previously studied cases.
The strong suppression of hydrogen Balmer lines in the spectra suggests interaction with a pre-existing circumstellar environment could be occurring at early times. In the SN\,2022wsp spectra, the absorption component of the \ion{Mg}{2} P~Cygni profile displays a double-trough feature on day $+$10 that disappears by day $+$20. The shape is well reproduced by the model without fine-tuning the parameters, suggesting that the secondary blueward dip is a metal transition that originates in the SN ejecta.
\end{abstract}

\keywords{supernovae: individual (SN\,2022wsp) --- ultraviolet: general --- techniques: spectroscopic}


\section{Introduction}\label{s:intro}
Type II supernovae (SNe) are characterized by the detection of hydrogen in their optical spectra. 
They can be differentiated photometrically by the shape of their light curve, which is primarily determined by the thickness of the hydrogen envelope of the exploding progenitor star.
An SN~II with a light curve that declines linearly in magnitude is classified as an SN~IIL, while an SN~II with a plateau lasting $\sim100$ days after the explosion is categorized as an SN~IIP. 
Nevertheless, recent studies suggest that the distinction between the SN~IIP and SN~IIL subtypes may not be convincing, as evidenced by continuum properties among the light curves of SNe~II  \citep{anderson_characterizing_2014, sanders_toward_2015, valenti_diversity_2016, rubin_unsupervised_2016}. The explosion mechanism producing Type II and a subset of Type I (Ib/Ic; stripped-envelope) SNe is widely accepted to be the core collapse of a star with zero-age main-sequence (ZAMS) mass $\geq 8$\,M$_{\odot}$; see \citet{filippenko_optical_1997} and \citet{gal-yam-2017} for reviews. However, the precise mechanism by which these explosions occur is not clear.

The spectral continuum of an SN~II peaks in the ultraviolet (UV) in the days and weeks following the explosion, and then continues to shift toward optical wavelengths over the next few months through a combination of cooling and line blanketing. 
Although the optical spectra of core-collapse SNe (CCSNe) have been extensively studied, the sample of UV spectra of CCSNe still remains scarce. 
This is in part due to the challenging nature of these observations, as they require rapid follow-up spectroscopy from space-based telescopes within three weeks after the SN explosion, before the radiation peak shifts from the UV to longer wavelengths. 
Moreover, the SNe must be sufficiently nearby to achieve a decent signal-to-noise ratio (S/N) in the UV. 
This is also complicated by the fact that the UV flux may be suppressed by dust attenuation along the SN-Earth line of sight. 
Despite these challenges, several programs have made significant strides in obtaining and analyzing UV spectra of SNe~II. 
UV spectroscopy has been carried out for only a handful of SNe~IIP, 
demonstrating a relatively high degree of uniformity in their UV (2000--3500\,\AA) 
spectroscopic properties at $\sim10$\,days after the explosion \citep{gal-yam_galex_2008, bufano_ultraviolet_2009, bayless_long-lived_2013, dhungana_extensive_2016}. 
Building on this research, \citet{vasylyev_early-time_2022} expanded the sample by including SN\,2021yja and demonstrating its resemblance to these H-rich CCSNe.

The UV spectra of SNe~IIP are also observed to have shared spectral line features. 
For example, SN~IIP UV spectra show the characteristic \ion{Mg}{2} $\lambda$2798 P~Cygni profile and several relatively broad emission ``bumps'' around 2200\,\AA, 2400\,\AA, and 2600\,\AA. These features can be attributed to a series of blended \ion{Fe}{2} and \ion{Ni}{2} lines \citep{brown_early_2007,gal-yam_galex_2008,bufano_ultraviolet_2009,dhungana_extensive_2016}. 
We emphasize that the sample size utilized in this study is limited and predominantly derived from a subset of SNe~II that have been observed through \textit{Hubble Space Telescope (HST)} and \textit{Swift}/UVOT programs. 
In order to gain a comprehensive understanding of the properties and behavior of their UV spectra, it is therefore imperative to construct a rather complete sample that allows further explorations of any commonality and diversity among SNe~II. Nevertheless, the results from these studies provide a crucial baseline to develop a comprehensive understanding of the nature of these commonly seen H-rich CCSNe.

Early-time UV spectra provide critical information about the dynamics and composition of the explosion since they are highly sensitive to the velocity of the expanding ejecta and the temperature.  Additionally, the line species and their strengths depict the metallicity and surface composition of the exploding star, thus tracing the pre-explosion behavior \citep{mazzali_applications_2000,dessart_quantitative_2005,dessart_quantitative_2006}.
Moreover, the UV flux is an excellent probe of the circumstellar environment of the progenitor, allowing one to identify additional heating sources such as interaction with the circumstellar medium \citep[CSM;][]{ben-ami_ultraviolet_2015}. 

Here we present two early-time UV spectra of a relatively nearby, young, and moderately reddened Type IIP SN\,2022wsp (DLT22q). 
The event was discovered on 02 Oct. 2022 at 23:59:19 (UTC dates are used throughout this paper) in the spiral galaxy NGC\,7448 by the Distance Less Than 40 (DLT40; \citealt{tartaglia_early_2018}) survey (see \citealp{bostroem_dlt40_2022}). 
The last nondetection was on 02 Oct. 03:51:46, with an upper limit of 19.6\,mag through a clear filter. Follow-up spectroscopy conducted on 05 Oct 23:46 suggested that SN\,2022wsp was an SN~II a few days after the explosion \citep{nagao_nuts_2022}. 
A ``preferred'' redshift of $z = 0.00732$ was reported by \citet{lu_h_1993},
and a distance of $25.04\pm 1.79$\,Mpc can be queried from the NASA/IPAC NED Database based on the cosmic microwave background (CMB) redshift and adopting a Hubble constant of H$_{0}=73$\,km\,s$^{-1}$\,Mpc$^{-1}$ \citep{riess_comprehensive_2022}. 

 We requested Target of Opportunity observations with {\it HST} to obtain UV and optical spectra of SN\,2022wsp (GO-16656; PI A. Filippenko).  
This {\it Letter}, which presents the UV and optical spectroscopy at relatively early phases, is organized as follows. Section~\ref{sec:obs_analysis} summarizes the {\it HST} observations. In Section~\ref{sec:analysis} we analyze the data based on \texttt{TARDIS} (a one-dimensional Monte-Carlo radiative-transfer code) fitting of the spectra. A summary of the study is given in Section~\ref{s:conc}. 

\section{Observations}~\label{sec:obs_analysis}
\subsection{HST/STIS}~\label{sec:hst_stis}
The explosion date of SN\,2022wsp was estimated as the midpoint between the last nondetection (MJD 59854.16) and the first detection (MJD 59855.00), which gives MJD 59854.58 $\pm$ 0.42. {\it HST} UV and optical follow-up spectroscopy of SN\,2022wsp was executed on 12 and 22 Oct. 2022, corresponding to $+$10 and $+$20 days after the explosion, respectively. 
Owing to technical and scheduling issues, the third epoch of observations was delayed until more than three months after the explosion. Unfortunately, by this time, the UV radiation of the SN had faded below the threshold required for proper spectral analysis; the peak of the SN spectral energy distribution had moved to optical wavelengths. The third epoch of our {\it HST} optical observations of SN\,2022wsp will be presented in future work, together with a comprehensive analysis of the photometric, spectroscopic, and polarimetric properties of the SN until the nebular phase (Vasylyev et al., in prep.).

\begin{figure*}
    \centering
    \includegraphics[width=1.0\textwidth]{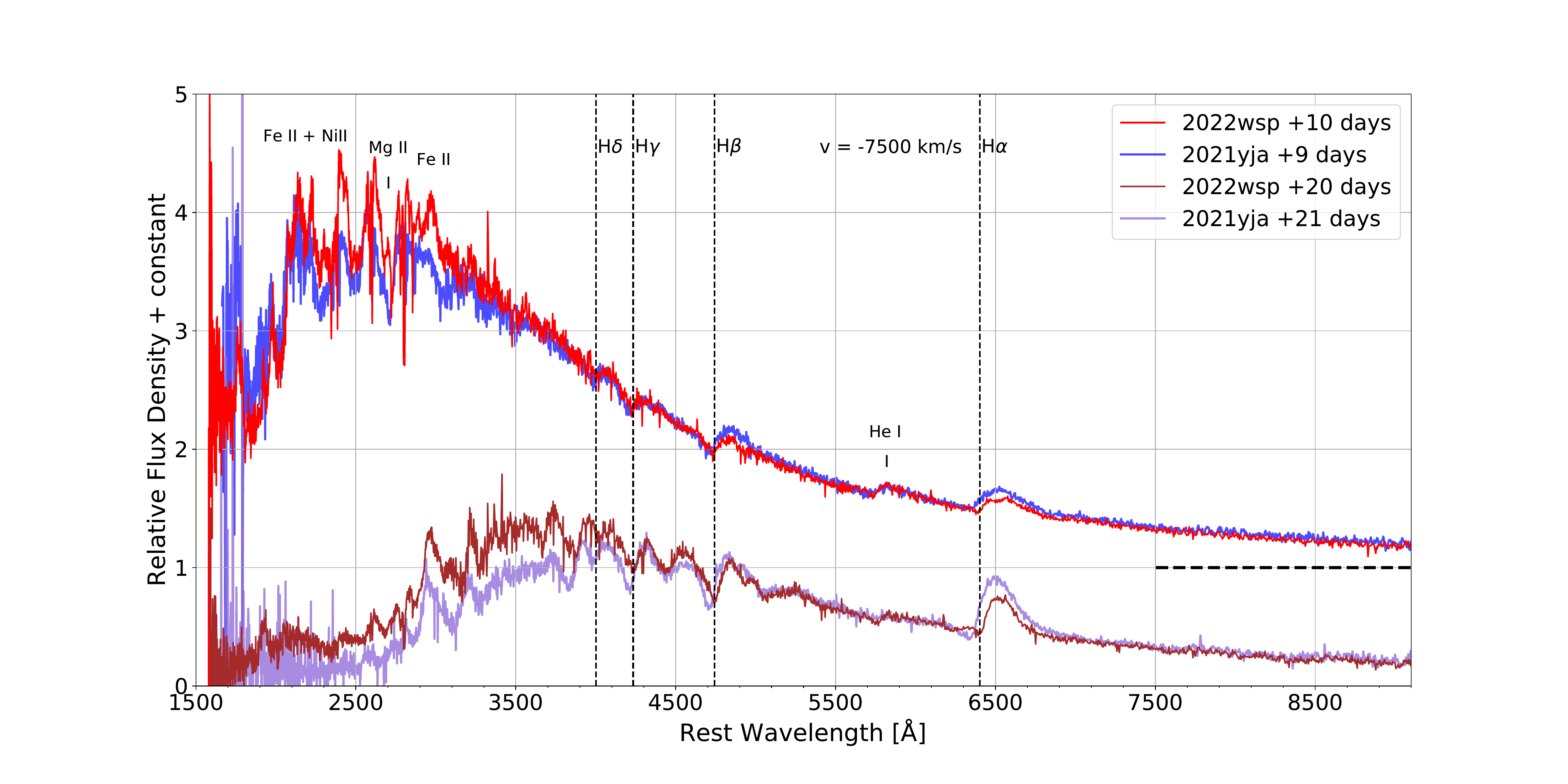}
    \caption{{\it HST}/STIS UV-optical spectra of SN\,2022wsp at +10 and +20 days compared with those of SN\,2021yja at similar phases.
    The SN\,2022wsp flux density has been scaled by the same constant for both days such that the continuum level red-ward of 7000 \AA matches that of SN\,2021yja, demonstrating the dramatic evolution and variation in the UV, while showing general agreement at optical wavelengths. The spectra have also been shifted arbitrarily by a constant for easier comparison.
    The wavelength scale has been corrected to the rest frame using the recession velocity of the host galaxy. Balmer lines at an expansion velocity of $v = 7500$\,km\,s$^{-1}$ are marked by vertical dotted lines. The zero flux level of the top two spectra is indicated by the horizontal dashed line
    }
    \label{fig:hst_uvo}
\end{figure*}

All spectroscopy was carried out using the CCD ($52^{\prime \prime} \times 52^{\prime \prime}$ field of view) and the Near-UV Multi-Anode MicroChannel Array (NUV-MAMA) detectors of the Space Telescope Imaging Spectrograph (STIS; \citealp{stis}).
For Epochs 1 and 2, observations of the mid-UV (MUV; 1570–3180\,\AA) with the G230L grating were made across six orbits of integration. 
One visit per epoch was used for the near-UV (NUV; 2900--5700\,\AA) and optical (5240--10270\,\AA) ranges with the G430L and G750L gratings, respectively. 
%
A detailed log of the {\it HST} observations can be found in Table \ref{Table:HSTobslog}. 
The UV-optical spectra at +10 and +20 days are presented in Figure~\ref{fig:hst_uvo}. 
All the {\it HST} data used in this paper can be found in MAST: \dataset[10.17909/0dxs-xd86]{http://dx.doi.org/10.17909/0dxs-xd86}.

\begin{figure}
    \centering
    \includegraphics[width=0.5\textwidth]{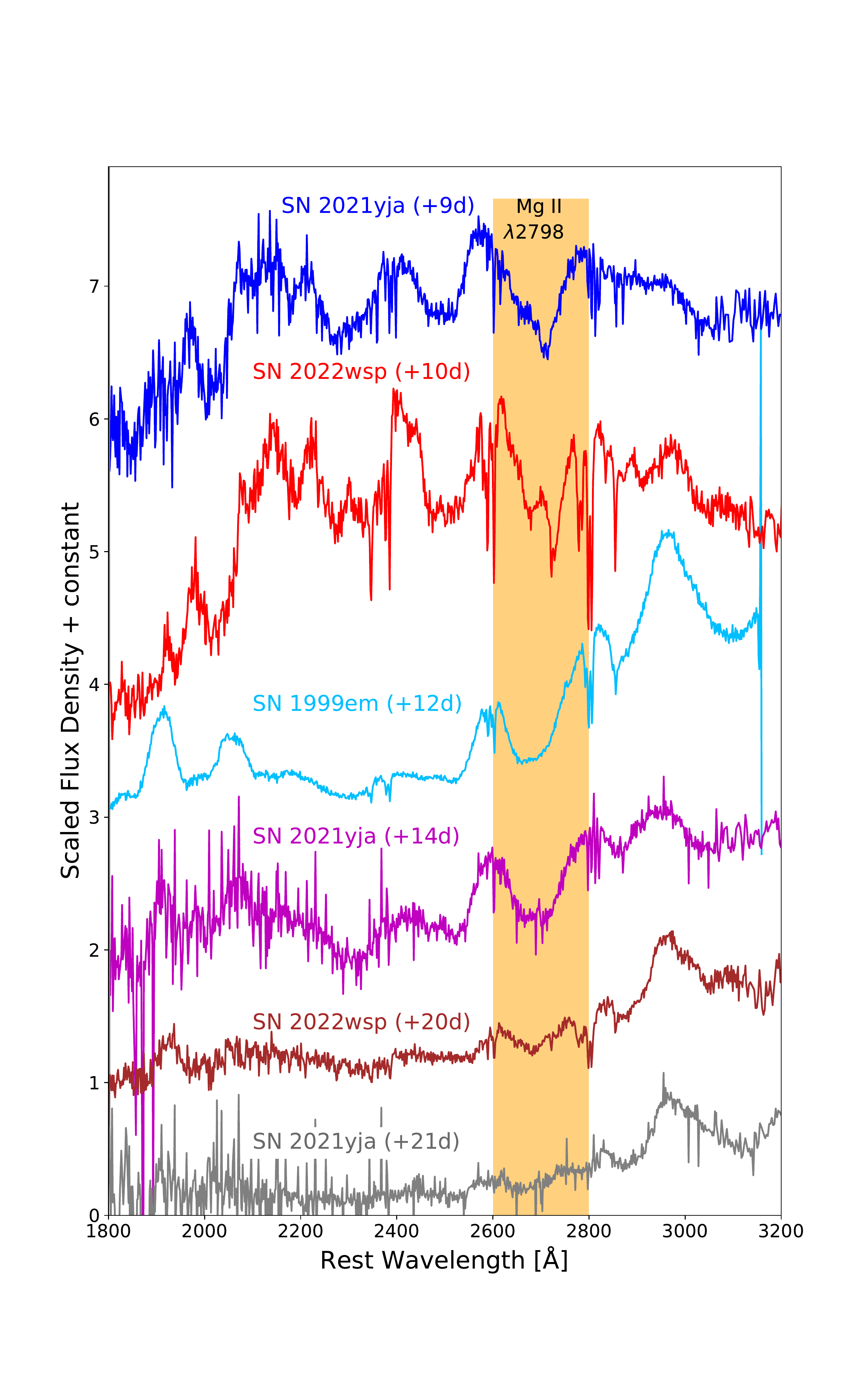}
    \caption{Zoomed-in view of the {\it HST}/STIS UV spectra of SN\,2022wsp at +10 and +20 days for clearer comparison to Type IIP SNe\,1999em \citep{baron_preliminary_2000} and 2021yja \citep{vasylyev_early-time_2022}. 
    }
    \label{fig:uv_comp}
\end{figure}



\section{Analysis and Discussion}~\label{sec:analysis}
In this section, we examine the spectra of SN\,2022wsp through a comparison with those of other well-sampled Type II and Type IIP SNe. This includes a discussion of the dust properties along the line of sight to the explosion and the modeling of the UV-optical spectra with \texttt{TARDIS}. 

\subsection{Dust Extinction}

The Galactic reddening along the line of sight to SN\,2022wsp gives $E(B-V)_{\text{MW}} = 0.05$\,mag, according to the extinction map derived by \citet{schlafly_measuring_2011}. 
We determined that the host galaxy also contributes significantly to the extinction ($E(B-V)_{\text{host}} = 0.3$\,mag), which results in a combined reddening value of $E(B-V)_{\text{tot}} = 0.35$\,mag (see Sec. \ref{s:tardis}). The relatively high reddening value of the SN is consistent with the explosion site being projected close to the nucleus of the host galaxy. 
The SN\,2022wsp spectra presented in this work are dereddened by the total extinction adopting an $R_{\text{V}} = 3.1$ dust law and the CCM89 model \citep{cardelli_relationship_1989} unless specified otherwise.  Also, all of the spectra have been corrected for the recession of the host galaxy NGC\,7448 using
$z = 0.00732$ (NED/IPAC Extragalactic Database23\footnote{See \url{https://ned.ipac.caltech.edu/}}; \citealt{lu_h_1993}).

\subsection{The Early-Time UV Spectrum}
\label{s:uv}
Here, we examine the UV spectral properties of SN\,2022wsp and compare them with those of SN\,2021yja as presented by \citet{vasylyev_early-time_2022} and with other SNe~IIP having high-S/N UV data. 

In Figure~\ref{fig:hst_uvo}, we present the {\it HST}/STIS UV-optical spectra of SN\,2022wsp obtained on days +10 and +20. 
For comparison, we also display the spectra of SN\,2021yja obtained on days +9 and +21.
Identifications of the major spectral features are provided by labels and vertical lines, indicating the presence of a broad Balmer series and the distinctive \ion{Mg}{2} $\lambda$2798 absorption line. 
These characteristics are also commonly observed in other SNe~II/IIP, such as SNe\,1999em, 2005cs, 2005ay, 2012aw, 2013ej, and 2021yja \citep{baron_preliminary_2000,brown_early_2007,gal-yam_galex_2008,bufano_ultraviolet_2009,bayless_long-lived_2013,vasylyev_early-time_2022}. 

The spectrum blueward of $\sim3000$\,\AA\ is dominated by a blend of \ion{Fe}{2}, \ion{Fe}{3}, and \ion{Ni}{3} lines, causing significant line blanketing \citep{dessart_quantitative_2005}. 
A series of broad emission peaks over this wavelength range can be attributed to regions of reduced line blanketing \citep{brown_early_2007}. 
As shown by the spectra of SNe\,2022wsp and 2021yja at day $+$20, the region between 2000\,\AA\ and 3000\,\AA\ appears to be significantly smoother compared to day +10. This can be understood as a consequence of a reinforced effect of line blanketing as the ejecta expand and cool over time.
Our observations provide further evidence of the general consistency in spectroscopic properties between SN\,2022wsp and other SNe~II/IIP, supporting the notion that they likely have similar progenitors and explosion mechanisms.

Figure~\ref{fig:uv_comp} compares the UV spectra of SN\,2022yja at +10 and +20 days with those of the Type IIP SNe\,1999em \citep{baron_preliminary_2000} and 2021yja \citep{vasylyev_early-time_2022}. While previous studies have suggested UV homogeneity among SN~IIP spectra, our data reveal both similarities and clear differences. These variations could result from a combination of factors, including differences in progenitor radius, explosion-date uncertainty, metallicity, density profile, photospheric temperature, velocity, reddening, and pre-explosion behavior.  

Figure~\ref{fig:hst_uvo} demonstrates a general agreement between SN\,2022wsp and SN\,2021yja at wavelengths longer than 3000\,\AA, around both +10 and +20 days.  
However, we observe some notable differences in the shape of the UV flux at wavelengths below 3000\,\AA\ at $\sim10$\,days, despite comparable observation epochs with tight constraints on the explosion date ($< 24$\,hr). Notably, the UV flux continuum of SN\,2022wsp appears to be elevated, with more pronounced emission and absorption features when compared to that of SN\,2021yja.

Modeling of the spectroscopic properties of SNe~II/IIP during the early photospheric phase using the non-local thermodynamic equilibrium (non-LTE) radiative transfer code \texttt{CMFGEN} \citep{hillier_treatment_1998} have been carried out in previous studies (see, e.g., SN\,1999em; \citealp{dessart_quantitative_2005}). The modeling addressed that the shape of the UV spectrum is highly sensitive to a set of physical parameters.
Specifically, a larger progenitor radius, a steeper radial density profile of the ejecta, a higher photospheric temperature, a lower metallicity, and a lower reddening can each produce a raised UV continuum, without significantly affecting the optical continuum. 
However, there exists a partial degeneracy between these parameters, particularly between reddening, photospheric temperature, and metallicity. A more detailed discussion of modeling parameters is presented in Section~\ref{s:tardis}. Our modeling of the early UV spectra of SN\,2022wsp demonstrates a general agreement with that of SN\,2021yja, although there may be plausible variations in the explosion properties. The differences in the spectra become less apparent around day +20.

As discussed by \cite{vasylyev_early-time_2022}, the \ion{Mg}{2} $\lambda$2798 line is of particular interest in the UV. Its shape has been used by previous studies to infer explosion properties of SNe~II, such as the origin of the UV emission \citep{brown_early_2007}. 
On day +10, SN\,2022wsp exhibits a distinct double-trough feature in the absorption component of \ion{Mg}{2} (see the 2600--2800\,\AA\ range in Figure \ref{fig:uv_comp}), 
which was not seen in any previous observations such as SNe\,1999em, 2005cs, 2005ay, 2012aw, 2013ej, and 2021yja \citep{baron_preliminary_2000,brown_early_2007,gal-yam_galex_2008,bufano_ultraviolet_2009,bayless_long-lived_2013,vasylyev_early-time_2022}. 

As discussed below in Section~\ref{s:tardis}, the double-trough \ion{Mg}{2} $\lambda$2798 profile can be well fitted by the one-dimensional (1D) radiative-transfer code, \texttt{TARDIS}. 
The surprisingly good reproduction suggests that the dip on the blue side is likely the result of overlapping \ion{Fe}{2} or \ion{Fe}{3} lines formed in a region close to the photosphere, rather than from an external source or developed from a complex geometry that may require further fine-tuning of the model parameters. 
The shape of this line changes rapidly over a few days, emphasizing the importance of performing observations over short time intervals. 
As shown by the bottom two spectra in Figure~\ref{fig:uv_comp}, the feature blueward of the \ion{Mg}{2} $\lambda$2798 absorption can be identified only at day +10 and has vanished by day +20. Other prominent features seen from SN\,2022wsp include the \ion{Fe}{2} line at 2900\,\AA\ and  \ion{Ti}{2} line blanketing around 3000\,\AA\ as labeled in Figure~\ref{fig:hst_uvo}.
These lines of singly ionized species were also identified in SNe\,2021yja \citep{vasylyev_early-time_2022} and 2005cs \citep{bufano_ultraviolet_2009}, and modeled in SN\,1999em \citep{dessart_quantitative_2005}. 


The day +10 spectrum of SN\,2022wsp presents a box-shaped flux excess around 2400\,\AA, which completely disappeared by day +20.
Interestingly, this feature is also evident in the \texttt{CMFGEN} model spectrum discussed by \citet{dessart_quantitative_2005} (see their Fig. 3) and \citet{dessart_quantitative_2006} (see their Figs. 10 and 11) for SN\,1999em on day +12, despite the observed spectrum displaying a relatively flat bump. 
It should be noted that this boxy feature resembles the absorption curve of \ion{Fe}{2} in the rectified spectra modeled for SN\,1999em near 2400\,\AA\ (see Fig. 3 of \citet{dessart_quantitative_2005}). The superposition of \ion{Fe}{2} together with \ion{Fe}{3} form the trough in the SN\,2021yja spectrum (day $+$9) near 2000\,\AA, which are also present in the $+$10 day spectrum of SN\,2022wsp.

We emphasize that the spectra presented in Figure~\ref{fig:uv_comp} are distinct from those of peculiar Type II SNe such as SN\,1987A, which display a sharp cutoff in the UV flux below 3000\,\AA\ \citep{kirshner_ultraviolet_1987,pun_ultraviolet_1995} by around 10 days after explosion. SN\,1987A was the explosion of a blue supergiant (BSG), rather than the red supergiant (RSG) progenitor of typical SNe~II/IIP. BSGs are expected to be more compact than RSGs and therefore cool more rapidly, increasing UV flux opacity due to line blanketing by metals at an earlier phase than SN~2022wsp. 
Additionally, other SNe~II that present clear evidence of ejecta-CSM interaction may show UV spectroscopic properties that are distinct from those of normal SNe~II/IIP. For example, a significantly blueshifted component in the \ion{Mg}{2} $\lambda$2798 P~Cygni profile and a rather smooth and elevated continuum below $\sim2600$\,\AA\ were reported for the Type IIb SN\,2013df \citep{ben-ami_ultraviolet_2015}.


\subsection{The Optical Spectrum}
Optical spectra of SN\,2022wsp were obtained at the same phases as our UV observations and are presented in Figure~\ref{fig:hst_uvo}.
As in the UV, the day +10 spectrum of SN\,2022wsp appears to be very similar to the day +9 spectrum of SN\,2021yja in the optical.
Both SNe can be described by the superposition of a relatively featureless continuum and a series of P~Cygni profiles of Balmer lines with blueshifted ($\geq 7000$\,km\,s$^{-1}$) absorption minima. 

There are, however, some subtle but important differences between the early-time spectra of SNe\,2022wsp and 2021yja.
For example, a suppressed, double-peaked H$\alpha$ profile can be identified in the day +10 spectrum of SN 2022wsp, which then develops into a more typical P-Cygni emission shape by day +20. 
By contrast, the SN 2021yja H$\alpha$ line resembles a more typical, rounded P Cygni profile. 
In Section~\ref{s:tardis}, we show that the fits fail to reproduce the strong suppression and the double-peaked shape, suggesting that interaction may be occurring with a pre-existing CSM.
One interpretation is that the ejecta interacting with CSM can effectively dampen the strength of the Balmer line profiles through the ``top-lighting'' effect \citep{branch_supernova_2000}.
The suppressed H$\alpha$ line blends together with \ion{He}{1} $\lambda$6678, forming a weak double-peaked, asymmetric emission feature. 
Meanwhile, the \ion{He}{1} $\lambda$5876 line is more defined in both epochs of SN\,2022wsp compared with SN\,2021yja.

\subsection{\texttt{TARDIS} Modeling of SN\,2022wsp}
\label{s:tardis}

Sophisticated radiative-transfer codes have been utilized to model SNe~II/IIP. 
For example, \citet{baron_type_2004} employed \texttt{PHOENIX}, while \citet{dessart_quantitative_2006} and  \citet{dessart_using_2008} made use of \texttt{CMFGEN} to perform detailed analyses of the physical properties of the explosion, such as temperature, density, and ionization structure. 
In order to extract the physical parameters of the explosion of SN\,2022wsp from its early spectroscopic evolution, we utilize a modified version of the Monte Carlo radiative-transfer code \texttt{TARDIS} repurposed for analyzing Type II SNe \citep{vogl_spectral_2019}; it was originally developed for SNe\,Ia \citep{kerzendorf_spectral_2014}. 

This code treats hydrogen excitation and ionization under non-LTE conditions, which is essential to accurately model the Balmer emission lines. 
We conduct a comparative analysis of SN\,2022wsp and that of SN\,2021yja, following the approach discussed in Section 3.5 of \cite{vasylyev_early-time_2022}. 
Similarly, we assumed a power-law density profile of the form $\rho = \rho_{0}(r/r_{0})^{-n}$, where $\rho_{0}$ denotes the density at a characteristic radius $r_{0}$ and $n$ represents the power-law index of the radial density profile. The calculation also assumes a homogeneous composition.
The models vary in the steepness of the density profile $n$, as well as the temperature, velocity, metallicity, and time since explosion. 

In addition to the parameters intrinsic to the SN explosion, we fit for different values of the host extinction, $E(B-V)_{\text{host}}$. 
The fitting is carried out using a machine-learning emulator, trained on a large grid of \texttt{TARDIS} simulations as described by \citet{vogl_spectral_2020}. We use the most up-to-date grid of models as described by \citet{csornyei_family_2023} (see their Table 2) for the training.

\begin{figure}
    \centering
    \includegraphics[width=0.5\textwidth]{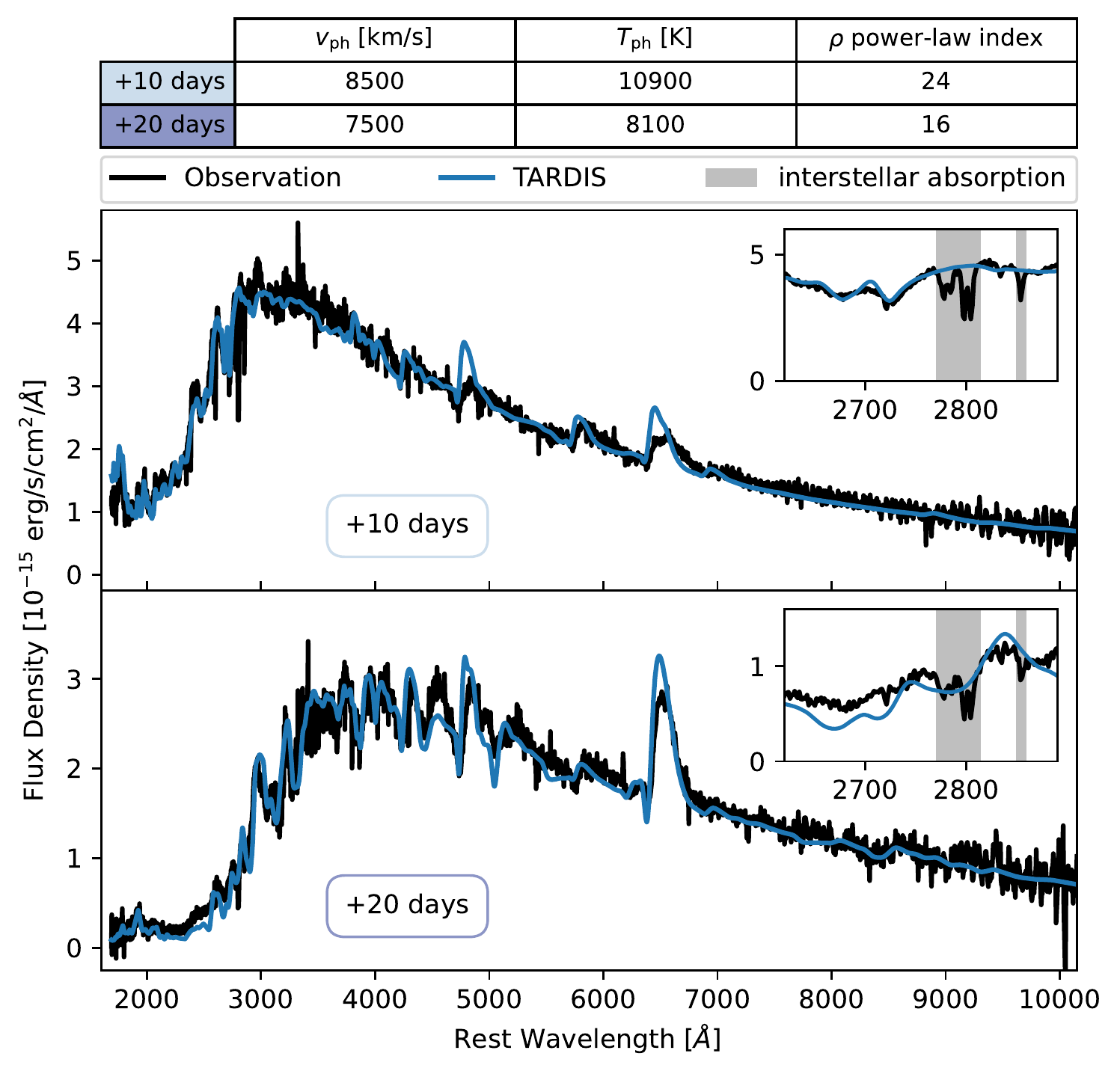}
    \caption{\texttt{TARDIS} fits to the {\it HST} STIS UV-optical spectra of SN\,2022wsp at days +10 (top) and +20 (bottom). The best-fit parameters are presented above the top panel. The embedded subpanels at upper right present a zoomed-in view of the \ion{Mg}{2} $\lambda$2798 feature. 
    The gray-shaded areas mark interstellar absorption lines.}
    \label{fig:tardis_fits}
\end{figure}

Figure~\ref{fig:tardis_fits} shows our \texttt{TARDIS} fits to the {\it HST} UV and optical spectra of SN\,2022wsp obtained on days +10 and +20. 
The plot includes a table containing the key fit parameters, namely the photospheric velocity $v_\mathrm{ph}$, the photospheric temperature $T_\mathrm{ph}$ (i.e., the gas temperature at the photosphere), and $n$. 
The best-fit model requires the presence of significant additional extinction from the host galaxy, with an estimated value of $E(B-V)_\mathrm{host}\approx 0.3$\,mag, as well as a supersolar metallicity. 
Given the uncertainties in the modeling, such as the approximate non-LTE treatment of metal species and the degeneracy between metallicity, temperature, and extinction, we provide only a qualitative estimate for the metallicity. 
In contrast to SN\,2021yja, the emission components of the Balmer P~Cygni profiles in the SN\,2022wsp spectra, particularly H$\alpha$, are not well fit by \texttt{TARDIS}. 
Such a discrepancy may be due to Balmer-line suppression by interaction with an ambient H-rich envelope. 

An embedded panel in Figure~\ref{fig:tardis_fits} presents a zoomed-in version on the \ion{Mg}{2}$\lambda$2798 absorption feature for both epochs.
Remarkably, the best-fit \texttt{TARDIS} model reproduces the double-trough profile that was initially discussed in Section~\ref{s:uv}. 
Such agreement between the observed spectrum and model suggests that the dip centered around 2675\,\AA\ is most likely an \ion{Fe}{2} transition formed within a part of the ejecta, rather than a high-velocity component from an external source or developed from a complex geometry that may require further fine-tuning of the model parameters. 
The narrow \ion{Mg}{2} $\lambda$$\lambda$2795, 2802 absorption doublets are excluded in the fitting procedure since they originate from the host galaxy and the Milky Way.


\begin{figure*}
    \centering
    \includegraphics[width=1\textwidth]{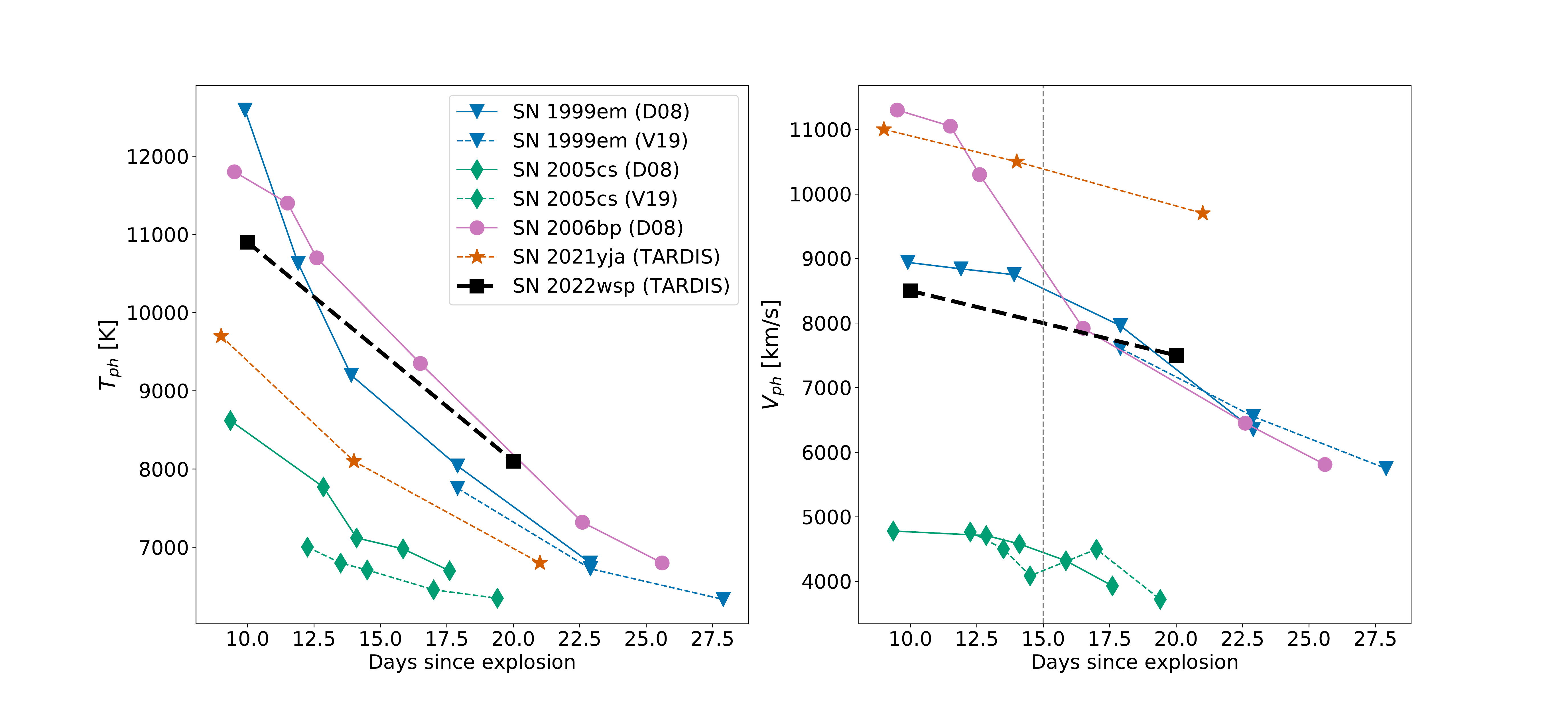}
    \caption{Temporal evolution of the photospheric temperature $T_{\rm ph}$ (left) and photospheric velocity $v_{\rm ph} $ (right) of SN\,2022wsp obtained from the \texttt{TARDIS} fit to its days +10 and +20 spectra compared to those of SNe\,1999em, 2005cs, 2006bp, and 2021yja  (\citealp{dessart_using_2008}, D08; \citealp{vogl_spectral_2019}, V19; \citealp{vasylyev_early-time_2022}). 
    Vertical dashed line indicates the velocity at days +15, $v_{\text{d}15}$, for comparison purposes.
    }
    \label{fig:modeling_comp}
\end{figure*}

\subsubsection{Evolution of the Photospheric Temperature and Velocity}
\label{sec:photo_temp_vel}

In Figure~\ref{fig:modeling_comp} we compare the photospheric temperature and velocity of SN\,2022wsp with those of other well-studied SNe~II/IIP. The physical parameters presented here are all derived based on modeling of the UV-to-optical spectra. The photospheric temperature of SN\,2022wsp at days +10 and +20 is 10,900\,K and 8,100\,K, respectively, both falling within the range typically found in SNe~IIP at similar phases. 
At early times ($\sim +12$\,days), the photospheric temperatures among these SNe span a wide range ($\sim 8500$--12,000\,K). 

The evolution of the temperature of SNe~IIP declines roughly linearly with time (at $> 10$\,days) until it  asymptotically approaches $\sim 6000$\,K, signaling the onset of the hydrogen recombination phase. 
The photospheric temperature is similar to that of SNe\,1999em and 2006bp, which is on the higher end of the range. SN\,2022wsp has a significantly higher inferred photospheric temperature than SNe\,2021yja and 2005cs.  
If all other fitting parameters were held constant for both SNe, the UV flux level is expected to be higher for SN\,2022wsp at a similar epoch owing to the higher $T_{\rm ph}$. 
As shown in Figure~\ref{fig:hst_uvo} and discussed in Section~\ref{s:uv}, the UV flux in the spectrum of SN\,2022wsp is elevated as compared with that of SN\,2021yja, consistent with the significantly higher $T_{\rm ph}$ inferred from our modeling. 

The photospheric velocity of SN\,2022wsp at days +10 and +20 gives 8500 and 7500\,km\,s$^{-1}$, respectively, as inferred from our \texttt{TARDIS} modeling. The velocity scale and evolution of SN\,2022wsp also agree with those derived from other SNe~II/IIP presented in Figure~\ref{fig:modeling_comp}.
The velocities of the H$\alpha$ absorption minima are consistent with those obtained from the \texttt{TARDIS} modeling. 
However, it is worth noting that for SNe~IIP, the \ion{Fe}{2} absorption minimum (e.g., $\lambda$5169) is the preferred method of estimating the photospheric velocity \citep{hamuy_distance_2001,leonard_distance_2002,dessart_quantitative_2005}. 
Moreover, the photospheric velocity at day +15 serves as a reliable indicator of the explosion energy \citep{dessart_determining_2010}. 
Based on the inferred photospheric velocities of SN\,2022wsp, which are comparable to those of SN\,1999em at +15\,days, it is likely that the two SNe have similar explosion energies.

\subsubsection{Radial Density Profile}
\label{sec:density}
Early-time observations of SNe~II suggest a steep density profile in the outer layers of the ejecta, which gradually flattens over time as the photosphere recedes into the inner layers \citep{dessart_using_2008,vogl_spectral_2019}. 
The inferred values of the radial density index $n$ are presented in the right-most column of Figure \ref{fig:tardis_fits}. 
In the case of SN\,2022wsp, a steep density profile is required to account for the strong suppression of the Balmer lines, especially H$\alpha$ and H$\beta$.

The power-law indices obtained from the fits for the first and second epochs are 24 and 16, respectively. The steep density profile suggests that the line-forming region is confined to a region close to the photosphere. 
It is worth noting that the density profile could have been altered by interaction, so it is uncertain whether a power law provides an appropriate description of the outer layers of the ejecta. The values for $n$ obtained for SN\,2022wsp are similar to those inferred from \texttt{TARDIS} modeling of SN\,2021yja \citep{vasylyev_early-time_2022}.

\section{Conclusions}\label{s:conc}

We present two epochs of {\it HST}/STIS UV-optical spectra on days +10 and +20 of the young, nearby, and relatively highly reddened Type IIP SN\,2022wsp. The UV spectrum of SN\,2022wsp is compared with that of previously studied SNe having high-S/N data at similar epochs.  Although SN\,2022wsp fits well within the framework of other well-studied SNe~IIP, there are a few key differences in the spectra. The \ion{Mg}{2} P~Cygni profile displays an unprecedented double-trough feature on day +10 that disappears by day +20. The origin of the blueward dip is most likely an overlapping \ion{Fe}{2} line.  
Overall, the differences in the spectra become less apparent around day $\sim 20$, highlighting the importance of conducting early-time observations in the UV to accurately constrain these parameters. Using the \texttt{TARDIS} code, the observed spectra were best fit by a photospheric velocity of 8500 (7500)\,km\,s$^{-1}$, a photospheric temperature of 10,900 (8100)\,K, a power-law index of 24 (16), and a supersolar metallicity on day +10 (20). The double-trough feature near the \ion{Mg}{2} absorption component is well fitted by this best-fit model. However, the suppressed emission components of H$\alpha$ and H$\beta$ are not represented by the \texttt{TARDIS} fit at day +10, suggesting that the outer layers of the ejecta may be interacting with CSM at these early phases. However, further investigation is needed to determine the validity of this interpretation. A follow-up paper to this work will present a detailed analysis of SN\,2022wsp optical spectroscopy, photometry, and spectropolarimetry (Vasylyev et al., in prep.).

\begin{deluxetable*}{lcccc}
\label{Table:HSTobslog}
    \tablecaption{HST Observation Log for SN\,2022wsp}
    \tablehead{ \colhead{Start Time [UTC]} & \colhead{Stop Time [UTC]} & \colhead{Exp. [s]}  & \colhead{Grating/Filter} & \colhead{$\lambda_{0}$ [\AA]}  }
    \startdata
        2022-10-12 10:30:55 & 2022-10-12 11:09:01 & 2286 & G230L & 2376 \\
        2022-10-12 11:57:50 & 2022-10-12 12:43:26 & 2736 & G230L & 2376\\
        2022-10-12 13:33:03 & 2022-10-12 14:18:39 & 2736 & G230L & 2376\\
        2022-10-12 15:20:08 & 2022-10-12 15:53:53 & 2025 & G230L & 2376\\
        2022-10-12 16:43:31 & 2022-10-12 17:29:07 & 2736 & G230L & 2376\\
        2022-10-12 18:18:46 & 2022-10-12 18:34:39 & 953 & G230L & 2376\\
        2022-10-12 18:38:15 & 2022-10-12 18:52:19 & 800 & G430L & 4300\\
        2022-10-12 18:56:28 & 2022-10-12 19:03:52 & 400 & G750L & 7751\\
        \hline
        2022-10-22 07:03:54	& 2022-10-22 07:42:00 & 2286 & G230L & 2376\\
        2022-10-22 08:30:50	& 2022-10-22 09:16:26 &	2736 & G230L & 2376\\
        2022-10-22 10:06:06	& 2022-10-22 10:51:42 &	2736 & G230L & 2376\\
        2022-10-22 11:53:12	& 2022-10-22 12:26:57 &	2025 & G230L & 2376\\
        2022-10-22 13:16:35	& 2022-10-22 14:02:11 &	2736 & G230L & 2376\\
        2022-10-22 14:51:51	& 2022-10-22 15:07:44 &	953 & G230L & 2376\\
        2022-10-22 15:11:20	& 2022-10-22 15:25:24 &	800 & G430L & 4300\\
        2022-10-22 15:29:33 & 2022-10-22 15:36:57 &	400 & G750L & 7551\\
        \hline
        2023-01-10 13:23:24 & 2023-01-10 14:01:30 & 2286 & G230L & 2376\\
        2023-01-10 14:50:14 & 2023-01-10 15:35:50 &	2736 & G230L & 2376\\
        2023-01-10 16:25:25 & 2023-01-10 17:11:01 &	2736 & G230L & 2376\\
        2023-01-10 18:12:28 & 2023-01-10 18:46:13 &	2025 & G230L & 2376\\
        2023-01-10 19:35:40 & 2023-01-10 20:02:49 &	1585 & G430L & 4300\\
        2023-01-10 20:06:58 & 2023-01-10 20:20:54 &	792 & G750L & 7551\\
    \enddata
\end{deluxetable*}


\begin{acknowledgments} 

This research was funded by {\it HST} grants AR-14259 and GO-16656 from the Space Telescope Science Institute (STScI), which is operated by the Association of Universities for Research in Astronomy (AURA), Inc., under NASA contract NAS5-26555. Additional generous financial support was provided to A.V.F.'s supernova group at U.C. Berkeley by Gary and Cynthia Bengier, Alan Eustace, Sunil Nagaraj, Steven Nelson, Landon Noll, Sandy Otellini, Christopher R. Redlich, Sanford Robertson, Clark and Sharon Winslow, Frank and Kathleen Wood, and other individual donors.
%
C.V. was supported for part of this work by the Excellence Cluster ORIGINS, which is funded by the Deutsche Forschungsgemeinschaft (DFG, German Research Foundation) under Germany's Excellence Strategy-EXC-2094-390783311.
G.S's research was supported through the Cal NERDS and UC LEADS programs.

This research made use of \texttt{TARDIS}, a community-developed software package for spectral
synthesis of SNe \citep{kerzendorf_spectral_2014, kerzendorf_wolfgang_2022_6299948}. The
development of \texttt{TARDIS} received support from GitHub, the Google Summer of Code
initiative, and ESA's Summer of Code in Space program. \texttt{TARDIS} is a fiscally
sponsored project of NumFOCUS. \texttt{TARDIS} makes extensive use of Astropy and Pyne.

\software{{Astropy \citep{astropy:2013, astropy:2018}, \texttt{TARDIS} \citep{kerzendorf_spectral_2014,vogl_spectral_2019}, uvotpy \citep{kuin_uvotpy_2014}, DAOPHOT \citep{stetson_daophot_1987}, IDL Astronomy user's library \citep{landsman_idl_1993}, SOUSA pipeline \citep{brown_sousa_2014}, Pyne \citep{scopatz_pyne_2012-1}}}
\bigskip

\end{acknowledgments}
\bigskip
\newpage
\clearpage
\bibliographystyle{aasjournal}
\bibliography{2022wsp}

\begin{thebibliography}{}
\expandafter\ifx\csname natexlab\endcsname\relax\def\natexlab#1{#1}\fi
\providecommand{\url}[1]{\href{#1}{#1}}

\bibitem[{Anderson {et~al.}(2014)Anderson, González-Gaitán, Hamuy,
  Gutiérrez, Stritzinger, E, Phillips, Schulze, Antezana, Bolt, Campillay,
  Castellón, Contreras, Jaeger, Folatelli, Förster, Freedman, González,
  Hsiao, Krzemiński, Krisciunas, Maza, McCarthy, Morrell, Persson, Roth,
  Salgado, Suntzeff, \& Thomas-Osip}]{anderson_characterizing_2014}
Anderson, J.~P., González-Gaitán, S., Hamuy, M., {et~al.} 2014, ApJ, 786, 67.
\newblock \url{https://doi.org/10.1088/0004-637x/786/1/67}

\bibitem[{{Astropy Collaboration} {et~al.}(2013){Astropy Collaboration},
  {Robitaille}, {Tollerud}, {Greenfield}, {Droettboom}, {Bray}, {Aldcroft},
  {Davis}, {Ginsburg}, {Price-Whelan}, {Kerzendorf}, {Conley}, {Crighton},
  {Barbary}, {Muna}, {Ferguson}, {Grollier}, {Parikh}, {Nair}, {Unther},
  {Deil}, {Woillez}, {Conseil}, {Kramer}, {Turner}, {Singer}, {Fox}, {Weaver},
  {Zabalza}, {Edwards}, {Azalee Bostroem}, {Burke}, {Casey}, {Crawford},
  {Dencheva}, {Ely}, {Jenness}, {Labrie}, {Lim}, {Pierfederici}, {Pontzen},
  {Ptak}, {Refsdal}, {Servillat}, \& {Streicher}}]{astropy:2013}
{Astropy Collaboration}, {Robitaille}, T.~P., {Tollerud}, E.~J., {et~al.} 2013,
  \aap, 558, A33

\bibitem[{{Astropy Collaboration} {et~al.}(2018){Astropy Collaboration},
  {Price-Whelan}, {Sip{\H{o}}cz}, {G{\"u}nther}, {Lim}, {Crawford}, {Conseil},
  {Shupe}, {Craig}, {Dencheva}, {Ginsburg}, {Vand erPlas}, {Bradley},
  {P{\'e}rez-Su{\'a}rez}, {de Val-Borro}, {Aldcroft}, {Cruz}, {Robitaille},
  {Tollerud}, {Ardelean}, {Babej}, {Bach}, {Bachetti}, {Bakanov}, {Bamford},
  {Barentsen}, {Barmby}, {Baumbach}, {Berry}, {Biscani}, {Boquien}, {Bostroem},
  {Bouma}, {Brammer}, {Bray}, {Breytenbach}, {Buddelmeijer}, {Burke},
  {Calderone}, {Cano Rodr{\'\i}guez}, {Cara}, {Cardoso}, {Cheedella}, {Copin},
  {Corrales}, {Crichton}, {D'Avella}, {Deil}, {Depagne}, {Dietrich}, {Donath},
  {Droettboom}, {Earl}, {Erben}, {Fabbro}, {Ferreira}, {Finethy}, {Fox},
  {Garrison}, {Gibbons}, {Goldstein}, {Gommers}, {Greco}, {Greenfield},
  {Groener}, {Grollier}, {Hagen}, {Hirst}, {Homeier}, {Horton}, {Hosseinzadeh},
  {Hu}, {Hunkeler}, {Ivezi{\'c}}, {Jain}, {Jenness}, {Kanarek}, {Kendrew},
  {Kern}, {Kerzendorf}, {Khvalko}, {King}, {Kirkby}, {Kulkarni}, {Kumar},
  {Lee}, {Lenz}, {Littlefair}, {Ma}, {Macleod}, {Mastropietro}, {McCully},
  {Montagnac}, {Morris}, {Mueller}, {Mumford}, {Muna}, {Murphy}, {Nelson},
  {Nguyen}, {Ninan}, {N{\"o}the}, {Ogaz}, {Oh}, {Parejko}, {Parley}, {Pascual},
  {Patil}, {Patil}, {Plunkett}, {Prochaska}, {Rastogi}, {Reddy Janga},
  {Sabater}, {Sakurikar}, {Seifert}, {Sherbert}, {Sherwood-Taylor}, {Shih},
  {Sick}, {Silbiger}, {Singanamalla}, {Singer}, {Sladen}, {Sooley},
  {Sornarajah}, {Streicher}, {Teuben}, {Thomas}, {Tremblay}, {Turner},
  {Terr{\'o}n}, {van Kerkwijk}, {de la Vega}, {Watkins}, {Weaver}, {Whitmore},
  {Woillez}, {Zabalza}, \& {Astropy Contributors}}]{astropy:2018}
{Astropy Collaboration}, {Price-Whelan}, A.~M., {Sip{\H{o}}cz}, B.~M., {et~al.}
  2018, \aj, 156, 123

\bibitem[{Baron {et~al.}(2004)Baron, Nugent, Branch, \&
  Hauschildt}]{baron_type_2004}
Baron, E., Nugent, P.~E., Branch, D., \& Hauschildt, P.~H. 2004, ApJ, 616, L91.
\newblock \url{https://iopscience.iop.org/article/10.1086/426506/meta}

\bibitem[{Baron {et~al.}(2000)Baron, Branch, Hauschildt, Filippenko, Kirshner,
  Challis, Jha, Chevalier, Fransson, Lundqvist, Garnavich, Leibundgut, McCray,
  Michael, Panagia, Phillips, Pun, Schmidt, Sonneborn, Suntzeff, Wang, \&
  Wheeler}]{baron_preliminary_2000}
Baron, E., Branch, D., Hauschildt, P.~H., {et~al.} 2000, ApJ, 545, 444.
\newblock \url{https://iopscience.iop.org/article/10.1086/317795/meta}

\bibitem[{Bayless {et~al.}(2013)Bayless, Pritchard, Roming, Kuin, Brown,
  Botticella, Dall'Ora, Frey, Even, Fryer, Maund, \&
  Fraser}]{bayless_long-lived_2013}
Bayless, A.~J., Pritchard, T.~A., Roming, P. W.~A., {et~al.} 2013, ApJ, 764,
  L13.
\newblock \url{https://doi.org/10.1088/2041-8205/764/1/l13}

\bibitem[{Ben-Ami {et~al.}(2015)Ben-Ami, Hachinger, Gal-Yam, Mazzali,
  Filippenko, Horesh, Matheson, Modjaz, Sauer, Silverman, Smith, \&
  Yaron}]{ben-ami_ultraviolet_2015}
Ben-Ami, S., Hachinger, S., Gal-Yam, A., {et~al.} 2015, ApJ, 803, 40.
\newblock \url{https://doi.org/10.1088/0004-637x/803/1/40}

\bibitem[{Bostroem {et~al.}(2022)Bostroem, Valenti, Sand, Wyatt, Lundquist,
  Andrews, Jencson, Dong, Janzen, Hosseinzadeh, Pearson, Meza, Shrestha, Hoang,
  \& Paraskeva}]{bostroem_dlt40_2022}
Bostroem, K.~A., Valenti, S., Sand, D.~J., {et~al.} 2022, Transient Name Server
  Discovery Report, 2022-2882, 1.
\newblock \url{https://ui.adsabs.harvard.edu/abs/2022TNSTR2882....1B}

\bibitem[{Branch {et~al.}(2000)Branch, Jeffery, Blaylock, \&
  Hatano}]{branch_supernova_2000}
Branch, D., Jeffery, D.~J., Blaylock, M., \& Hatano, K. 2000, Publications of
  the Astronomical Society of the Pacific, 112, 217, publisher: IOP Publishing.
\newblock \url{https://iopscience.iop.org/article/10.1086/316510/meta}

\bibitem[{Brown {et~al.}(2014)Brown, Breeveld, Holland, Kuin, \&
  Pritchard}]{brown_sousa_2014}
Brown, P.~J., Breeveld, A.~A., Holland, S., Kuin, P., \& Pritchard, T. 2014,
  Astrophysics and Space Science, 354, 89.
\newblock \url{https://doi.org/10.1007/s10509-014-2059-8}

\bibitem[{Brown {et~al.}(2007)Brown, Dessart, Holland, Immler, Landsman,
  Blondin, Blustin, Breeveld, Dewangan, Gehrels, Hutchins, Kirshner, Mason,
  Mazzali, Milne, Modjaz, \& Roming}]{brown_early_2007}
Brown, P.~J., Dessart, L., Holland, S.~T., {et~al.} 2007, ApJ, 659, 1488.
\newblock \url{https://doi.org/10.1086/511968}

\bibitem[{Bufano {et~al.}(2009)Bufano, Immler, Turatto, Landsman, Brown,
  Benetti, Cappellaro, Holland, Mazzali, Milne, Panagia, Pian, Roming,
  Zampieri, Breeveld, \& Gehrels}]{bufano_ultraviolet_2009}
Bufano, F., Immler, S., Turatto, M., {et~al.} 2009, ApJ, 700, 1456.
\newblock \url{https://doi.org/10.1088/0004-637x/700/2/1456}

\bibitem[{Cardelli {et~al.}(1989)Cardelli, Clayton, \&
  Mathis}]{cardelli_relationship_1989}
Cardelli, J.~A., Clayton, G.~C., \& Mathis, J.~S. 1989, ApJ, 345, 245, aDS
  Bibcode: 1989ApJ...345..245C.
\newblock \url{https://ui.adsabs.harvard.edu/abs/1989ApJ...345..245C}

\bibitem[{{Cs\"ornyei, G.} {et~al.}(2023){Cs\"ornyei, G.}, {Vogl, C.},
  {Taubenberger, S.}, {Fl\"ors, A.}, {Blondin, S.}, {Cudmani, M. G.}, {Holas,
  A.}, {Kressierer, S.}, {Leibundgut, B.}, \& {Hillebrandt,
  W.}}]{csornyei_family_2023}
{Cs\"ornyei, G.}, {Vogl, C.}, {Taubenberger, S.}, {et~al.} 2023, A\&A, 672,
  A129.
\newblock \url{https://doi.org/10.1051/0004-6361/202245379}

\bibitem[{Dessart \& Hillier(2005)}]{dessart_quantitative_2005}
Dessart, L., \& Hillier, D.~J. 2005, A\&A, 437, 667.
\newblock \url{https://ui.adsabs.harvard.edu/abs/2005A&A...437..667D/abstract}

\bibitem[{Dessart \& Hillier(2006)}]{dessart_quantitative_2006}
---. 2006, A\&A, 447, 691.
\newblock \url{http://arxiv.org/abs/astro-ph/0510526}

\bibitem[{Dessart {et~al.}(2010)Dessart, Livne, \&
  Waldman}]{dessart_determining_2010}
Dessart, L., Livne, E., \& Waldman, R. 2010, MNRAS, 408, 827.
\newblock \url{https://doi.org/10.1111/j.1365-2966.2010.17190.x}

\bibitem[{Dessart {et~al.}(2008)Dessart, Blondin, Brown, Hicken, Hillier,
  Holland, Immler, Kirshner, Milne, Modjaz, \& Roming}]{dessart_using_2008}
Dessart, L., Blondin, S., Brown, P.~J., {et~al.} 2008, ApJ, 675, 644.
\newblock \url{https://iopscience.iop.org/article/10.1086/526451}

\bibitem[{Dhungana {et~al.}(2016)Dhungana, Kehoe, Vinko, Silverman, Wheeler,
  Marion, Zheng, Fox, Akerlof, Biro, Borkovits, Cenko, Clubb, Filippenko,
  Ferrante, Gibson, Graham, Hegedus, Kelly, Kelemen, Lee, Marschalko, Molnár,
  Nagy, Ordasi, Pal, Sarneczky, Shivvers, Szakats, Szalai, Szegedi-Elek,
  Székely, Szing, Takáts, \& Vida}]{dhungana_extensive_2016}
Dhungana, G., Kehoe, R., Vinko, J., {et~al.} 2016, ApJ, 822, 6.
\newblock \url{http://arxiv.org/abs/1509.01721}

\bibitem[{Filippenko(1997)}]{filippenko_optical_1997}
Filippenko, A.~V. 1997, Annual Review of Astronomy and Astrophysics, 35, 309.
\newblock \url{https://doi.org/10.1146/annurev.astro.35.1.309}

\bibitem[{Gal-Yam(2017)}]{gal-yam-2017}
Gal-Yam, A. 2017, in Alsabti A. W., Murdin P., eds, Handbook of Supernovae,
  Springer International Publishing, Cham,195

\bibitem[{Gal-Yam {et~al.}(2008)Gal-Yam, Bufano, Barlow, Baron, Benetti,
  Cappellaro, Challis, Ellis, Filippenko, Foley, Fox, Hicken, Kirshner,
  Leonard, Li, Maoz, Matheson, Mazzali, Modjaz, Nomoto, Ofek, Simon, Small,
  Smith, Turatto, Dyk, \& Zampieri}]{gal-yam_galex_2008}
Gal-Yam, A., Bufano, F., Barlow, T.~A., {et~al.} 2008, ApJ, 685, L117.
\newblock \url{https://iopscience.iop.org/article/10.1086/592744/meta}

\bibitem[{Hamuy {et~al.}(2001)Hamuy, Pinto, Maza, Suntzeff, Phillips, Eastman,
  Smith, Corbally, Burstein, Li, Ivanov, Moro-Martin, Strolger, Souza, Anjos,
  Green, Pickering, Gonzalez, Antezana, Wischnjewsky, Galaz, Roth, Persson, \&
  Schommer}]{hamuy_distance_2001}
Hamuy, M., Pinto, P.~A., Maza, J., {et~al.} 2001, ApJ, 558, 615.
\newblock \url{https://doi.org/10.1086/322450}

\bibitem[{Hillier(1998)}]{hillier_treatment_1998}
Hillier, D.~J. 1998, ApJ, 496, 407.
\newblock \url{https://doi.org/10.1086/305350}

\bibitem[{Kerzendorf {et~al.}(2022)Kerzendorf, Sim, Vogl, Williamson, Pássaro,
  Flörs, Camacho, Jančauskas, Harpole, Nöbauer, Lietzau, Mishin, Tsamis,
  Boyle, Shingles, Gupta, Desai, Klauser, Beaujean, Suban-Loewen, Heringer,
  Barna, Gautam, Fullard, Cawley, Singhal, Smith, Barbosa, Sondhi, Patel,
  Varanasi, Gillanders, Arya, O'Brien, Eweis, Reinecke, Bylund, Bentil, Savel,
  Yu, Eguren, Alam, Magee, Livneh, Rajagopalan, Mishra, Reichenbach, Jain,
  Floers, Brar, Singh, Talegaonkar, Kowalski, Selsing, Sofiatti, Aggarwal,
  Sarafina, Patra, Singh~Rathore, Patel, Sharma, Gupta, Wahi, Sandler, Volodin,
  Dasgupta, Yap, Kharkar, Nayak~U, Kolliboyina, \&
  Kumar}]{kerzendorf_wolfgang_2022_6299948}
Kerzendorf, W., Sim, S., Vogl, C., {et~al.} 2022, tardis-sn/tardis: TARDIS
  v2022.2.27, vrelease-2022.2.27,  Zenodo, doi:10.5281/zenodo.6299948.
\newblock \url{https://doi.org/10.5281/zenodo.6299948}

\bibitem[{Kerzendorf \& Sim(2014)}]{kerzendorf_spectral_2014}
Kerzendorf, W.~E., \& Sim, S.~A. 2014, MNRAS, 440, 387.
\newblock \url{https://doi.org/10.1093/mnras/stu055}

\bibitem[{Kirshner {et~al.}(1987)Kirshner, Sonneborn, Crenshaw, \&
  Nassiopoulos}]{kirshner_ultraviolet_1987}
Kirshner, R.~P., Sonneborn, G., Crenshaw, D.~M., \& Nassiopoulos, G.~E. 1987,
  ApJ, 320, 602.
\newblock \url{https://ui.adsabs.harvard.edu/abs/1987ApJ...320..602K}

\bibitem[{Kuin(2014)}]{kuin_uvotpy_2014}
Kuin, P. 2014, Astrophysics Source Code Library, ascl:1410.004.
\newblock \url{https://ui.adsabs.harvard.edu/abs/2014ascl.soft10004K}

\bibitem[{Landsman(1993)}]{landsman_idl_1993}
Landsman, W.~B. 1993, 52, 246.
\newblock \url{https://ui.adsabs.harvard.edu/abs/1993ASPC...52..246L}

\bibitem[{Leonard {et~al.}(2002)Leonard, Filippenko, Gates, Li, Eastman, Barth,
  Bus, Chornock, Coil, Frink, Grady, Harris, Malkan, Matheson, Quirrenbach, \&
  Treffers}]{leonard_distance_2002}
Leonard, D.~C., Filippenko, A.~V., Gates, E.~L., {et~al.} 2002, Publications of
  the Astronomical Society of the Pacific, 114, 35.
\newblock \url{https://doi.org/10.1086/324785}

\bibitem[{Lu {et~al.}(1993)Lu, Hoffman, Groff, Roos, \& Lamphier}]{lu_h_1993}
Lu, N.~Y., Hoffman, G.~L., Groff, T., Roos, T., \& Lamphier, C. 1993, ApJ
  Supplement Series, 88, 383, aDS Bibcode: 1993ApJS...88..383L.
\newblock \url{https://ui.adsabs.harvard.edu/abs/1993ApJS...88..383L}

\bibitem[{Mazzali(2000)}]{mazzali_applications_2000}
Mazzali, P.~A. 2000, A\&A, 363, 705.
\newblock \url{https://ui.adsabs.harvard.edu/abs/2000A&A...363..705M/abstract}

\bibitem[{Nagao {et~al.}(2022)Nagao, Pastorello, \& Reguitti}]{nagao_nuts_2022}
Nagao, T., Pastorello, A., \& Reguitti, A. 2022, Transient Name Server
  Classification Report, 2022-2910, 1.
\newblock \url{https://ui.adsabs.harvard.edu/abs/2022TNSCR2910....1N}

\bibitem[{Prichard {et~al.}(2022)Prichard, Welty, \& Jones}]{stis}
Prichard, L., Welty, D., \& Jones, A. 2022, Baltimore: STScI

\bibitem[{Pun {et~al.}(1995)Pun, Kirshner, Sonneborn, Challis, Nassiopoulos,
  Arquilla, Crenshaw, Shrader, Teays, Cassatella, Gilmozzi, Talavera,
  Wamsteker, Fransson, \& Panagia}]{pun_ultraviolet_1995}
Pun, C. S.~J., Kirshner, R.~P., Sonneborn, G., {et~al.} 1995, ApJ Supplement
  Series, 99, 223.
\newblock \url{http://adsabs.harvard.edu/doi/10.1086/192185}

\bibitem[{Riess {et~al.}(2022)Riess, Yuan, Macri, Scolnic, Brout, Casertano,
  Jones, Murakami, Anand, Breuval, Brink, Filippenko, Hoffmann, Jha, Kenworthy,
  Mackenty, Stahl, \& Zheng}]{riess_comprehensive_2022}
Riess, A.~G., Yuan, W., Macri, L.~M., {et~al.} 2022, ApJ Letters, 934, L7,
  publisher: The American Astronomical Society.
\newblock \url{https://dx.doi.org/10.3847/2041-8213/ac5c5b}

\bibitem[{Rubin \& Gal-Yam(2016)}]{rubin_unsupervised_2016}
Rubin, A., \& Gal-Yam, A. 2016, ApJ, 828, 111, publisher: The American
  Astronomical Society.
\newblock \url{https://dx.doi.org/10.3847/0004-637X/828/2/111}

\bibitem[{Sanders {et~al.}(2015)Sanders, Soderberg, Gezari, Betancourt,
  Chornock, Berger, Foley, Challis, Drout, Kirshner, Lunnan, Marion, Margutti,
  McKinnon, Milisavljevic, Narayan, Rest, Kankare, Mattila, Smartt, Huber,
  Burgett, Draper, Hodapp, Kaiser, Kudritzki, Magnier, Metcalfe, Morgan, Price,
  Tonry, Wainscoat, \& Waters}]{sanders_toward_2015}
Sanders, N.~E., Soderberg, A.~M., Gezari, S., {et~al.} 2015, ApJ, 799, 208.
\newblock \url{https://dx.doi.org/10.1088/0004-637X/799/2/208}

\bibitem[{Schlafly \& Finkbeiner(2011)}]{schlafly_measuring_2011}
Schlafly, E.~F., \& Finkbeiner, D.~P. 2011, ApJ, 737, 103.
\newblock \url{https://iopscience.iop.org/article/10.1088/0004-637X/737/2/103}

\bibitem[{Scopatz {et~al.}(2012)Scopatz, Romano, Wilson, \&
  Huff}]{scopatz_pyne_2012-1}
Scopatz, A.~M., Romano, P.~K., Wilson, P.~P., \& Huff, K.~D. 2012, Transactions
  of the American Nuclear Society, 107, 985.
\newblock
  \url{http://www.scopus.com/inward/record.url?scp=84876547920&partnerID=8YFLogxK}

\bibitem[{Stetson(1987)}]{stetson_daophot_1987}
Stetson, P.~B. 1987, Publications of the Astronomical Society of the Pacific,
  99, 191.
\newblock \url{https://ui.adsabs.harvard.edu/abs/1987PASP...99..191S}

\bibitem[{Tartaglia {et~al.}(2018)Tartaglia, Sand, Valenti, Wyatt, Anderson,
  Arcavi, Ashall, Botticella, Cartier, Chen, Cikota, Coulter, Valle, Foley,
  Gal-Yam, Galbany, Gall, Haislip, Harmanen, Hosseinzadeh, Howell, Hsiao,
  Inserra, Jha, Kankare, Kilpatrick, Kouprianov, Kuncarayakti, Maccarone,
  Maguire, Mattila, Mazzali, McCully, Melandri, Morrell, Phillips, Pignata,
  Piro, Prentice, Reichart, Rojas-Bravo, Smartt, Smith, Sollerman, Stritzinger,
  Sullivan, Taddia, \& Young}]{tartaglia_early_2018}
Tartaglia, L., Sand, D.~J., Valenti, S., {et~al.} 2018, ApJ, 853, 62.
\newblock \url{https://dx.doi.org/10.3847/1538-4357/aaa014}

\bibitem[{Valenti {et~al.}(2016)Valenti, Howell, Stritzinger, Graham,
  Hosseinzadeh, Arcavi, Bildsten, Jerkstrand, McCully, Pastorello, Piro, Sand,
  Smartt, Terreran, Baltay, Benetti, Brown, Filippenko, Fraser, Rabinowitz,
  Sullivan, \& Yuan}]{valenti_diversity_2016}
Valenti, S., Howell, D.~A., Stritzinger, M.~D., {et~al.} 2016, MNRAS, 459,
  3939.
\newblock \url{https://ui.adsabs.harvard.edu/abs/2016MNRAS.459.3939V}

\bibitem[{Vasylyev {et~al.}(2022)Vasylyev, Filippenko, Vogl, Brink, Brown,
  Jaeger, Matheson, Gal-Yam, Mazzali, Modjaz, Patra, Rowe, Smith, Dyk,
  Williamson, Yang, Zheng, deGraw, Fox, Gates, Jennings, \&
  Rich}]{vasylyev_early-time_2022}
Vasylyev, S.~S., Filippenko, A.~V., Vogl, C., {et~al.} 2022, ApJ, 934, 134,
  publisher: The American Astronomical Society.
\newblock \url{https://dx.doi.org/10.3847/1538-4357/ac7220}

\bibitem[{Vogl {et~al.}(2020)Vogl, Kerzendorf, Sim, Noebauer, Lietzau, \&
  Hillebrandt}]{vogl_spectral_2020}
Vogl, C., Kerzendorf, W.~E., Sim, S.~A., {et~al.} 2020, A\&A, 633, A88.
\newblock \url{http://arxiv.org/abs/1911.04444}

\bibitem[{Vogl {et~al.}(2019)Vogl, Sim, Noebauer, Kerzendorf, \&
  Hillebrandt}]{vogl_spectral_2019}
Vogl, C., Sim, S.~A., Noebauer, U.~M., Kerzendorf, W.~E., \& Hillebrandt, W.
  2019, A\&A, 621, A29.
\newblock \url{https://ui.adsabs.harvard.edu/abs/2019A&A...621A..29V/abstract}

\end{thebibliography}

\listofchanges
\end{document}